\documentclass[preprint,12pt]{elsarticle}

\usepackage{amssymb}

\usepackage{amsmath}
\usepackage{algpseudocode}
\usepackage{algorithm}

\def \N {\mathbb{N}}
\def \Z {\mathbb{Z}}

\def \R {\mathbb{R}}

\def \w {\mathbf{w}}

\def \x {\mathbf{x}}

\def \m {\mathbf{m}}

\def \a {\mathbf{a}}

\def \J {\mathbf{J}}

\def \E {\mathbb{E}}

\newtheorem{defi}{Definition}[section]

\newtheorem{thm}{Theorem}[section]

\journal{ }
\begin{document}
\begin{frontmatter}

\title{A local Echo State Property through the largest Lyapunov exponent}

\author{Gilles Wainrib}
%\cortext[cor1]{Ecole Normale Superieure, Departement d'Informatique, Paris, France.\ead{gilles.wainrib@ens.fr}}
%\cortext[cor2]{NeuroMathComp, Inria Sophia; UNIC, CNRS Gif; Minds, Jacobs University Bremen.\ead{mathieu.galtier@gmx.com}}

\address{Ecole Normale Superieure, Departement d'Informatique, Paris, France.}

\author{Mathieu N. Galtier}

\address{NeuroMathComp, Inria Sophia; UNIC, CNRS Gif; Minds, Jacobs University Bremen.}

\begin{abstract}
Echo State Networks are efficient time-series predictors, which highly depend on the value of the spectral radius of the reservoir connectivity matrix. Based on recent results on the mean field theory of driven random recurrent neural networks, enabling the computation of the largest Lyapunov exponent of an ESN, we develop a cheap algorithm to establish a local and operational version of the Echo State Property.
\end{abstract}


\end{frontmatter}

\section{Introduction}
Echo State Networks (ESN) are neural networks designed for performing complex non-linear regression or classification tasks, such as non-linear time-series forecasting \cite{jaeger2001short, jaeger2004harnessing}. As an instance of a more general framework called reservoir computing \cite{lukovsevivcius2009survey}, the ESN architecture is based on a randomly connected recurrent neural network, called reservoir, which is driven by a temporal input. The state of the reservoir is a rich representation of the history of the inputs \cite{buonomano1995temporal}, so that a simple linear combination of the reservoir neurons is often a good predictor of the future of the inputs. The computation of the output connections can be done explicitly and corresponds to the minimization of the relative entropy between the network and the inputs dynamics \cite{galtier2014relative}, for which the associated gradient descent may be implemented with biologically plausible learning rules \cite{galtier2013biological}.

In this paper, we focus on the input-driven reservoir, which may be governed by a variety of dynamical systems beyond random neural networks \cite{dambre2012information}, provided they produce consistent reservoir dynamics for a given input. This condition is of primary importance since its violation systematically leads to irrelevant results. In the original paper \cite{jaeger2001short}, Jaeger has given a condition, which he names \textit{Echo State Property} (ESP), guaranteeing that the network states are consistent. This definition of the ESP and the equivalent formulations manipulate left infinite input time-series assuming that the initial condition occurs at $t=-\infty$. If $n$ is the number of neurons in the reservoir, $\x(t) \in \R^n$ is the state of the reservoir at time $t\in \Z$ and $u(t) \in \R$ is the input to the reservoir of time $t\in \Z$. The ESP definition can be summarized as
\begin{defi}[ESP \cite{jaeger2001short}]
	A network has the ESP if the network state $\x(t)$ is uniquely determined by any left-infinite input sequence $\{u(t-s) : s \in \N\}$.
	\label{def: ESP}
\end{defi}
In other words, it means that the initial condition of the network (at $t = -\infty$) does not influence the trajectory of the states, which corresponds to the property that the input-driven network has a unique global attractor \cite{cheban2004global}. The ESP seems to be important in practice to design efficient reservoirs. Indeed, a network without ESP would have a poor accuracy in the inevitable presence of perturbations or noise: a small perturbation could bring the network to states it has never seen before, destroying the prediction capabilities of the network. Put differently, the network has to have some fading memory so that the initial conditions and perturbations do not impact the accuracy in the long term.

A fundamental result is that a bound on the maximum singular value $\eta$ of the network connectivity matrix $\J \in \R^{n \times n}$ can provide the global ESP for every input. More specifically, if the dynamics of the network is governed by
\begin{equation}
	\x_{i}(t+1) = S\left(\sum_{j=1}^n  \J_{ij} \x_j(t) + \m_i u(t)\right):= G_i(\x(t),t)
\end{equation}
where $\m \in \R^n$ is the input matrix, and $S(.)$ is a sigmoid function with unit slope at the origin, then the following result holds:
\begin{thm}[\cite{jaeger2001short}]
	If $\eta<1$, then the global ESP holds for every input.
	\label{thm: sing val}
\end{thm}
%\begin{proof}
%	Consider
%	\begin{equation}
%		\x_i(t+1)= S\left(\sigma \sum_{j=1}^n  \J_{ij} \x_j(t) + \m_i \u_t\right)\mbox{ with initial condition }\x(0)=\x_0
%	\end{equation}
%
%	\begin{equation}
%		\x'_i(t+1)= S\left(\sigma \sum_{j=1}^n  \J_{ij} \x'_j(t) + \m_i \u_t\right)\mbox{ with initial condition }\x'(0)=\x'_0
%	\end{equation}
%	Then, denoting $a_i(t)=\sigma \sum_{j=1}^n  \J_{ij} \x_j(t) + \m_i \u_t$,  $a'_i(t)=\sigma \sum_{j=1}^n  \J_{ij} \x'_j(t) + \m_i \u_t$ and $\delta(t):=a_i(t)-a'_i(t)$, we have
%	\begin{equation}
%		||a(t+1)-a'(t+1)||^2 = ||J.\delta(t)||^2 = \delta(t)^T .(J^T.J). \delta(t) \leq \eta(J) ||\delta(t)||^2
%	\end{equation}
%	where $\eta(J)$ is the largest singular value of $J$. Therefore, assuming $S(.)$ is $1$-Lipschitz, we have:
%	\begin{equation}
%		||x(t+1)-x'(t+1)||^2 \leq  \eta(J) K^2||x(t)-x'(t)||^2
%	\end{equation}
%	We conclude that if $\eta(J) < 1$, then all the trajectories starting from different initial conditions converge to same attractor.
%\end{proof}

It is important to observe that the sufficient condition in \ref{thm: sing val} holds for the largest singular value $\eta$ and not for the largest eigenvalue modulus $\rho$ (also called spectral radius), which are different for most matrices. Indeed, as pointed out in \cite{zhang2012nonlinear}, the theory of random matrices gives a relationship between the maximum singular value $\eta$ and the maximum eigenvalue $\rho$ of the random matrix $\J$ when the number of neurons tends to infinity. First, using recent results on the empirical spectral distribution of random matrices \cite{tao2010random}, one can show that large random matrices, whose entries are i.i.d. random variables with mean $0$, finite variance $\frac{\sigma^2}{\sqrt{n}}$ , have eigenvalues which tend to cover uniformly the disk of radius $\sigma$ as the number of neurons tends to infinity. For these matrices, the non-scaled standard deviation of the weights $\sigma$ is in fact equal to the spectral radius $\rho$. Second, one can use results concerning the right edge of the Marchenko-Pastur convergence \cite{marvcenko1967distribution, geman1980limit, bai2010spectral} to show that $\eta \to 2 \sigma$ when the number of neurons tends to infinity. From this result, as mentioned in \cite{zhang2012nonlinear}, it is clear that the condition on the singular values translates to
\begin{thm}
When the number of neurons tends to infinity (and with the appropriate scaling of the weights variance by $\frac{1}{\sqrt{n}}$) the ESP holds for all inputs if $\rho = \sigma<1/2$.
\end{thm}
Interestingly, there is here a clear gap between the theoretical sufficient condition $\eta<1$ (i.e $\sigma<1/2$) and the condition $\rho < 1$ (i.e $\sigma < 1$) which seems to be valid in practice \cite{lukovsevivcius2012practical}. Based on the notion of structured singular value and on concepts from control theory \cite{lohmiller1998contraction}, a tighter sufficient condition has been derived involving the computation of the infimum of the maximal singular values of the connectivity matrix for variety of underlying norms \cite{buehner2006tighter}. Despite its improvement over the classical singular value, this criterion is difficult to compute in practice, remains poorly understood from the point of view of random matrix theory, and does not respond to the problem of finding a criterion which depends on input, as we will discuss below. It is also interesting to mention the recent work \cite{zhang2012nonlinear}, where the concentration of measure phenomenon \cite{ledoux2005concentration} is used to prove that:
\begin{thm}[\cite{zhang2012nonlinear}]
	If $\rho < 1-\epsilon$, then for any 
	$\x,\tilde{\x} \in \R^n$, the probability that $||G(\x,t)-G(\tilde{\x},t)||>||\x-\tilde{\x}||$ is exponentially small when the number of neurons is large.
\label{thm: sigma < 1}
\end{thm}
This result may seem sufficient to prove the contraction property with high probability, implying the ESP when $\sigma<1$ with high probability. Actually, one must be careful because this result does not imply that
$\mathbb{P}[\forall \x,\tilde{\x} \in \R^n,\forall t>0, ||G(\x,t)-G(\tilde{\x},t)||>||\x-\tilde{\x}||]$ is small with high probability, which is a much stronger result. However, the authors claim that their result shows why choosing $\sigma$ close but smaller than one is sufficient in practice. In a sense, they argue that networks which do not verify criterion of Theorem \ref{thm: sing val} can still perform well in applications.

On the other side, it is also instructive to look for a necessary condition for the ESP. When the spectral radius $\rho$ is larger than one, then the trivial null equilibrium of the system with zero input is linearly unstable, and Jaeger has shown that:
\begin{thm}[\cite{jaeger2001short}]
When $\rho > 1$, the ESP does not hold for the null input.
\end{thm}
This result is in fact related to the existence of chaotic attractors as shown in \cite{sompolinsky1988chaos}. Therefore, there is no hope for an ESP for all inputs beyond $\rho = 1$. However, in practice \cite{lukovsevivcius2012practical}, it may be important to increase $\rho$ above $1$ to improve the ESN performance (to increase the memory for instance). If we want to go beyond $\rho = 1$, we need to drop the requirement to have the ESP for all inputs. It has recently been argued that one can define an ESP with respect to a particular input (or a set of inputs) \cite{manjunath2013echo}. Intuitively, this means that a network driven by an input will not display excessive irregularity if it has the ESP with respect to that input. In \cite{manjunath2013echo}, a bound for the ESP is also provided
\begin{thm}[\cite{manjunath2013echo}]
If $\underset{j \to \infty}{\mbox{lim sup}} \sum_{i=-1}^{-j}\Big(C_i - (1+\ln(2))\Big) I(C_i>2) >\frac{\ln(\|\J\|)}{2}$, with $C_i$ the smallest absolute component of the vector $\m u(i)$ and $I$ is the indicator function, then the network has the ESP with respect to $u$.
\end{thm}
Intuitively, this bound plays with the saturation of the sigmoid and will be efficient if the inputs are strong enough to drive the network in the saturating regime. Although this is a loose bound, it has the interesting property that the network may have temporarily non-contracting dynamics and still have the ESP. These ideas are clearly related to the fact that stimulating a chaotic system can result in a synchronized non-chaotic response, as shown in the context of random neural networks in \cite{rajan2010stimulus}.

In this paper, we aim at contributing to the debate about the ESP using a mean-field approach applied to non-autonomous random neural networks in the large $n$ limit. This theory derives a self-consistent statistical description of the reservoir dynamics unravelling the transition between regularity and irregularity in the network, based on a Lyapunov stability analysis.
Although brought very recently into the field of echo-state networks by \cite{massar2013mean}, this theoretical approach has a long history, dating back to early works on spin-glass models \cite{sompolinsky1981dynamic,sompolinsky1982relaxational}, followed by applications to random neural networks dynamics as in \cite{sompolinsky1988chaos, cessac1994mean, PhysRevLett.69.3717, faugeras2009constructive}. The rigorous justification of this heuristic approach is non-trivial and has been resolved by \cite{arous1995large,moynot2002large,cabana2013large} using large deviations techniques. These mathematical results actually requires to add an (arbitrary) small white-noise perturbation to the reservoir dynamics, in order to be able to use a change of probability formula (e.g. Girsanov Theorem) which is at the heart of the large deviation proof. The rigorous proof of the mean-field equations when this additional noise is removed remains open to our knowledge, but this is not a real problem in the ESN framework since adding such noise term is actually used in practice as a form of regularization, shown to be equivalent to the classical Tikhonov regularization \cite{bishop1995training}.

The network we consider in this paper is a leaky integrator ESN \cite{jaeger2007optimization} defined over a regular graph with degree $ \alpha n$, proportional to $n$. This means that every neuron in the network is only connected to $\alpha n$ other neurons, which is often used in practice to reduce computational complexity. To apply the mean-field theory, we will assume that $n$ goes to infinity, but consider $\alpha \in (0,1]$ to be a constant. The connections between neurons are weighted: we write $\J_{ij}$ the weight from neuron $j$ to neuron $i$. The weights are independent random variables satisfying:
\begin{equation*}
\E(\J_{ij}) = 0 \quad \mbox{and} \quad \E(\J_{ij}^2)= \frac{\sigma^2}{n} <+\infty
\end{equation*}
This quenched hypothesis excludes any dynamics on the weights: they are kept constant after having been randomly drawn.

Given a one-dimensional input time series $u:\{1\cdots T\} \to \R$, the classical neural network discrete dynamics is
\begin{equation}
	\x_{i}(t+1) = (1-l \tau)\x_i(t) + \tau S\left(\sum_{j\to i}  \J_{ij} \x_j(t) + \m_i u(t)\right)
	\label{eq: system}
\end{equation}
where $\x(t) \in \R^n$ corresponds to the activity of all the neurons in the network at time $t$.  The vector of feedforward connections $\m \in \R^n$ is made of i.i.d. random variables satisfying $\E(\m_i)=0$, $\E(\m_i^2) = m^2$. The numbers $l$ and $\tau$ are in $[0,1]$ and control the timescale of the ESN dynamics. The function $S(.)$ is a typical odd sigmoid with $S(0)=0$, $S'(0)=1$, $S'(x)>0$ and $x S''(x)\leq 0$. Note that it implies it is a 1-Lipschitz function. Actually, the following computations become explicit when a particular choice is made: $S(x)=\text{erf}(\frac{\sqrt{\pi}}{2} x)$ (which follows the requirements above). We write $\displaystyle \sum_{j \to i}$ the summation of incoming information to a neuron which is only done over the neurons which are connected (through the graph) to the considered neuron. 

The paper is organized as follows: in section \ref{sec: mean field}, we derive a mean field theory of driven leaky integrator recurrent neural networks (RNNs) on a regular graph, and we show how it can be used to find the frontier between order and disorder for the network dynamics. Then, in section \ref{sec: local ESP} we show how this can be used to define a computable condition guaranteeing an operational version of the ESP.

\section{Mean-field theory for leaky ESN on regular graphs}\label{sec: mean field}

\subsection{Mean-field equations}
From the seminal work \cite{sompolinsky1988chaos}, recently extended to the framework of stimulus driven RNN \cite{rajan2010stimulus, massar2013mean}, one can derive a self-consistent equation describing the statistical properties of the reservoir activity in the large $n$ limit, which is known as the mean-field theory. In this section, we present an extension of \cite{massar2013mean} to leaky RNNs on regular graphs.

The key idea is to make the assumption that the variables $\x_i(t)$ are i.i.d. and independent of $\J$ and $\m$. This makes possible to use the central limit theorem on $\sum_{j\to i} \J_{ij} \x_j(t)$ which can thus be considered as a Gaussian process. When $k = \alpha n \to + \infty$, all the $\a_i(t)=\sum_{j\to i} \J_{ij} \x_j(t) + \m_i u(t)$ for $i \in \{1..n\}$ tend to behave as centered Gaussian variables with variance 
\begin{equation*}
a^2(t) = \E[\a_i(t)^2] = \alpha \sigma^2 \gamma^2(t) + m^2 u(t)^2
\end{equation*}
where $\gamma^2(t)$ denotes the variance of $\x_i(t)$ (independent of $i$). The iteration equation $\x_i(t+1)=(1- l \tau) \x_i(t) + \tau S\big(\a_i(t)\big)$ is going to help us derive the mean-field dynamical system describing the variance of the $\x_i$. However, the independence between $\x_i(t)$ and $S(\a_i(t))$ is not granted and we cannot simply add their variance. Nonetheless, we can compute
%\begin{equation}
%\begin{array}{rcl}
%	\gamma^2(t+1)  & = &\E[\x_i(t+1)^2] \\
%	& = & (1- l \tau)^2 \gamma^2(t) + \tau^2 F\big(a^2(t)\big) \\
%	& + & \tau(1-l \tau) R(t,t)
%\end{array}
%	\label{eq: consistency variances}
%\end{equation}
\begin{equation}
	\gamma^2(t+1) = (1- l \tau)^2 \gamma^2(t) + \tau^2 F\big(a^2(t)\big) + 2\tau(1-l \tau) R(t,t)
	\label{eq: consistency variances}
\end{equation}

%\begin{equation}
%	\gamma^2(t+1) = (1- l \tau)^2 \gamma^2(t) + \tau^2 F\left(\alpha \sigma^2 \gamma^2(t) + m^2 u(t)^2\right)
%	\label{eq: consistency variances}
%\end{equation}
with \begin{equation}
	F(z^2) = (2\pi)^{-1/2}\int_\R S^2(zx)e^{-x^2/2}dx=  \frac{2}{\pi}\arcsin\left(\frac{\pi z^2}{2+\pi z^2}\right)
	\label{eq: F}
\end{equation}
according to the technical result in the appendix of \cite{williams1998computation}, and 
\begin{equation}
	R(s,t) = \E[\x_i(s)S\big(\a_i(t)\big)]= (1 - l \tau) R(s-1,t) + \tau Q(s-1,t)
	\label{eq: R}
\end{equation}
where
\begin{equation}
Q(s,t) = \E\big[S\big(\a_i(s)\big)S\big(\a_i(t)\big)\big]
\end{equation}
Using again the result in \cite{williams1998computation}, we can show that
\begin{multline}
Q(s,t) = G\Big(C(s,t), \gamma^2(s), \gamma^2(t) \Big)\\
= \frac{2}{\pi} \sin^{-1}\left(\frac{\pi}{2} \frac{\alpha \sigma^2 C(s,t) + m^2 u(s) u(t)}{\sqrt{\big(1 + \frac{\pi}{2}a^2(s)\big)\big(1 + \frac{\pi}{2} a^2(t)}\big)} \right)
\end{multline} 
where
\begin{equation}
C(s,t) = \E[\x_i(s) \x_i(t)] = (1-l \tau) C(s,t-1) + \tau R(s,t-1)
\label{eq: C}
\end{equation} 

The recursive combination of equations \eqref{eq: consistency variances}, \eqref{eq: R} and \eqref{eq: C} provides a consistent description of the global variance of the neurons. An algorithm is provided in algorithm \ref{alg: lambda(sigma)}.

\subsection{Order-disorder transition}\label{sec: order - disorder}
The consistency equation \eqref{eq: consistency variances} characterizes the transition between order and disorder in the network as a function of the variance of the connections $\sigma^2$ and the sparsity coefficient $\alpha$. We first illustrate this phenomenon in the autonomous case and then discuss its impact in the input driven case.

\subsubsection{Without input} 
the terms $\x_i(t)$ and $S(\a_i(t))$ are independent, and the third term in \eqref{eq: consistency variances} disappears. Thus, let us study the autonomous dynamical system $\gamma^2(t+1) = (1- l \tau)^2\gamma^2(t) + \tau^2 F\big(\alpha \sigma^2 \gamma^2(t)\big)$. Due to the properties of the sigmoid function $S$, the function $F$ is increasing, concave and satisfies $F(0)=0$ and $F'(0)=1$. Therefore, the function $\Psi:x \mapsto (1- l \tau)^2 x + \tau^2 F(\alpha \sigma^2 x)$ is also increasing and concave. Therefore, the slope at $0$, denoted $\mu=\Psi'(0)$, is the effective parameter controlling the phase transition, and is given by 
\begin{equation}
	\mu = (1- l \tau)^2 + \tau^2 \alpha \sigma^2
\end{equation}
This leads to a simple characterization of the behavior of the system for different values of $\mu$:
\begin{itemize}
	\item $\gamma^2(t)$ converges to $\gamma^2_{\infty}=0$ if $\mu < 1$
	\item $\gamma^2(t)$ converges to a limit value $\gamma^2_{\infty}>0$ if $\mu > 1$
\end{itemize}
In the first situation $\sigma<\sigma^*=\sqrt{\frac{l}{\alpha}(\frac{2}{\tau} - l)}$, all neuron variables converge to the quiescent state, whereas the network behavior becomes irregular as soon as $\sigma > \sigma^*$. Note that this generalizes the classical results of \cite{sompolinsky1988chaos,cessac1994mean} dealing with the case $\tau = \alpha = l = 1$, a case which is also treated in \cite{hermans2012recurrent}, where stability criteria are established for dynamical systems defining recurrent kernels for infinite-dimensional ESN.

\subsubsection{With inputs, largest Lyapunov exponent}
When the system is driven by external inputs, the network will never go to a quiescent state. Indeed, it is clear from equation \eqref{eq: consistency variances} that the situation $\gamma^2(t)=0$ will never happen. But one should not conclude that the network is always disordered because it could be strongly locked to the inputs, which is another way of defining the notion of order in such systems. The network will be said to be in order (resp. disorder) when a small perturbation independent of the inputs will vanish (rep. grow) with time. This corresponds to the notion of Lyapunov stability for the input driven system. The largest Lyapunov is below 1 in the case of robustness of the dynamics to small perturbations (order), and above 1 when the dynamics is significantly impacted by small perturbations, as is the case in chaotic systems (disorder). Formally, the largest lyapunov exponent can be defined as:
\begin{equation}
\lambda[u]:=\lim_{t\to \infty,\delta(0)\to 0} \left(\frac{\delta^2(t)}{\delta^2(0)}\right)^{1/t}
\label{eq: lyapu def}
\end{equation}
where $\delta(t)$ is a distance at time $t$ between two trajectories of \eqref{eq: system} starting with different initial conditions separated by $\delta(0)$. More precisely, let us define $\delta(t)$ such that $\x_i(t)-\x'_i(t) \sim \mathcal{N}$ where $\x_i$ and $\x'_i$ are two solutions of \eqref{eq: system} starting from two different initial conditions with $\x_i(0)-\x'_i(0) \sim \mathcal{N}(0,\delta(0)^2)$. In the situation where $\delta(t)$ is small, we have the following recurrence equation:
%\begin{eqnarray*}
%	\a_i(t+1)-\a'_i(t+1)&=& \sum_{j\to i}\J_{ij} \left(S\big(\a_j(t)\big)-S\big(\a'_j(t)\big)\right)\\
%	&=& \sum_{j\to i}\J_{ij}S'\left(\frac{\a_j(t) + \a'_j(t)}{2}\right)(\a_j(t) - \a'_j(t)) + o\big(\delta(t)\big)
%\end{eqnarray*}
\begin{equation*}
\begin{array}{rcl}
&&\x_i(t+1)-\x'_i(t+1)\\
&=& (1-l\tau) \big(\x_i(t)-\x'_i(t) \big)  + \tau \left(S\big(\a_i(t)\big) - S\big(\a_i'(t)\big) \right)\\
&=& (1-l\tau) \big(\x_i(t)-\x'_i(t) \big)  \\ &+& \tau S'\big(\a_i(t)\big)\left(\sum_{j\to i}\J_{ij} \big(\x_j(t) - \x'_j(t)\big) \right) + o\big(\delta(t)\big)\\
\end{array}
\end{equation*}
Therefore, one obtains the following relationship on the variances:
\begin{equation}
\begin{array}{rcl}
	\delta^2(t+1) &=& (1-l\tau)^2 \delta^2(t) \\
	& + & \tau^2 \alpha \sigma^2 \Phi\left(\alpha\sigma^2\gamma^2(t)+ m^2 u(t)^2 \right)\delta^2(t) \\
	&+&o(\delta^2(t))
\end{array}
	\label{eq: lyapu recurrence}
\end{equation}
with 
\begin{equation}
	\Phi(z^2) := (2\pi)^{-1/2}\int S'^2(z x)e^{-x^2/2}dx = \frac{1}{\sqrt{1 + \pi z^2}}.
	\label{eq: phi}
\end{equation}
When $\gamma^2(t)$ is obtained by solving iteratively \eqref{eq: consistency variances}, one can find the local Lyapunov exponent:
\begin{equation}
	\lambda(t):=(1-l\tau)^2 + \tau^2 \alpha \sigma^2 \Phi\left(\alpha\sigma^2\gamma^2(t)+ m^2 u(t)^2 \right)
	\label{eq:local}
\end{equation}
When $\lambda(t)[u]<1$, local asymptotic stability is ensured and the reservoir tends to be synchronized by the input, whereas when $\lambda(t)[u]>1$, small perturbations are exponentially amplified and the reservoir is likely to enter a chaotic regime. It is natural that this measure depends on time because, for instance in the case $\sigma>1$, synchronized states will only appear during periods when the input is sufficiently large compared to $\sigma$.

Combining \eqref{eq: lyapu def} and \eqref{eq: lyapu recurrence}, one can define a global finite horizon largest Lyapunov exponent as:
\begin{equation}
	\lambda_T[u]:=\left(\prod_{t=1}^T \lambda(t) \right)^{\frac{1}{T}}
	\label{eq: lambda T}
\end{equation}
where $\lambda(t)$ is defined in \eqref{eq:local}. 
Furthermore, at this stage, one already obtains an important property, showing that adding external input can only stabilize the system. Indeed, since $\Phi \leq 1$ (due to the fact that $|S'|\leq 1$), we always have the following inequality:
\begin{equation}
	\lambda_T \leq \mu
	\label{eq:inequality}
\end{equation}
Therefore, if the system without external input is in the ordered phase, namely when $\mu<1$, then it is also in the ordered phase ($\lambda_T[u]<1$) for all input. This results supports the fact that, in practice, $\rho<1$ is a sufficient condition for the ESP.

\begin{figure*}[t]
\centering
\includegraphics[width=0.4\textwidth]{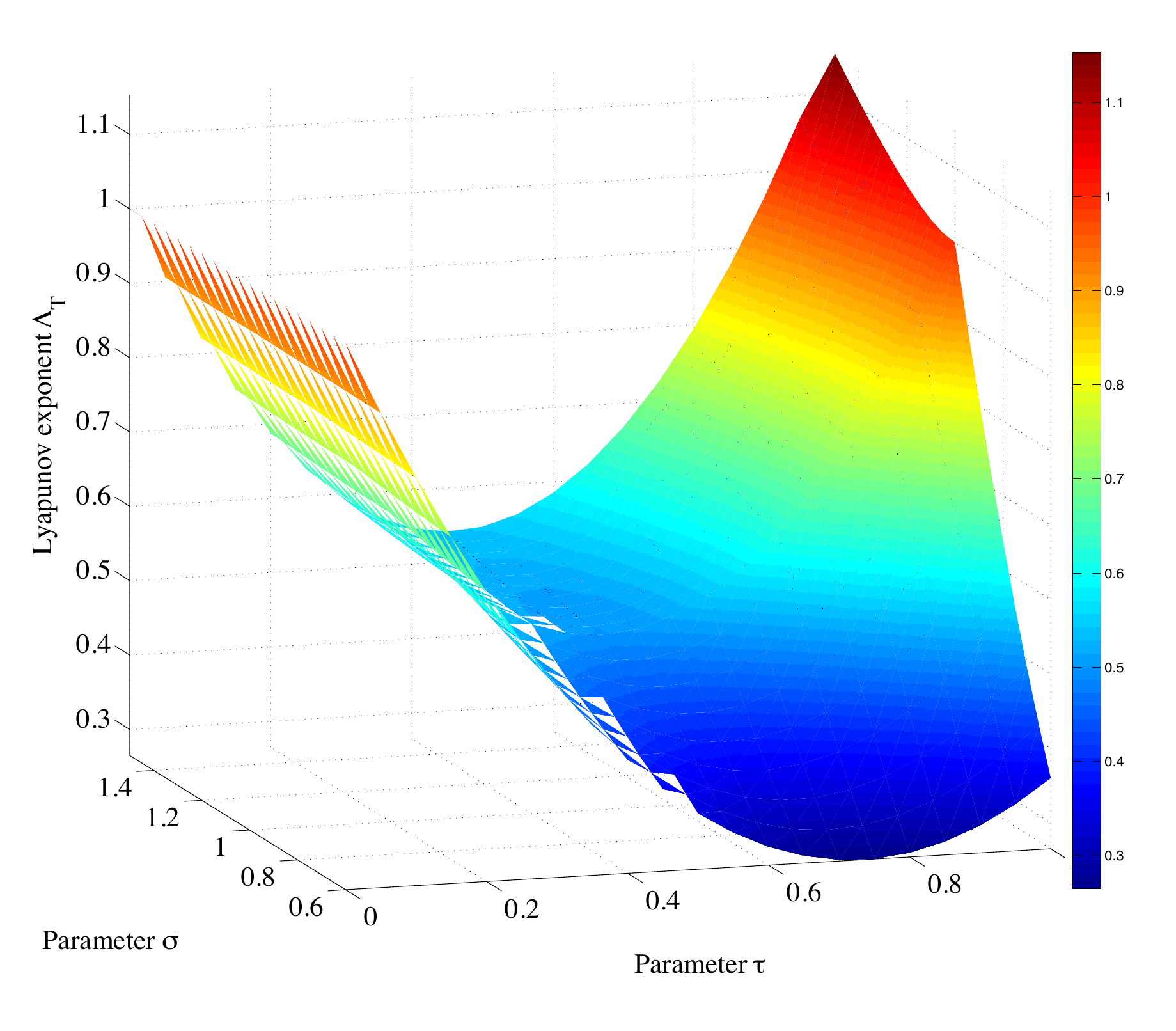}
\includegraphics[width=0.43\textwidth]{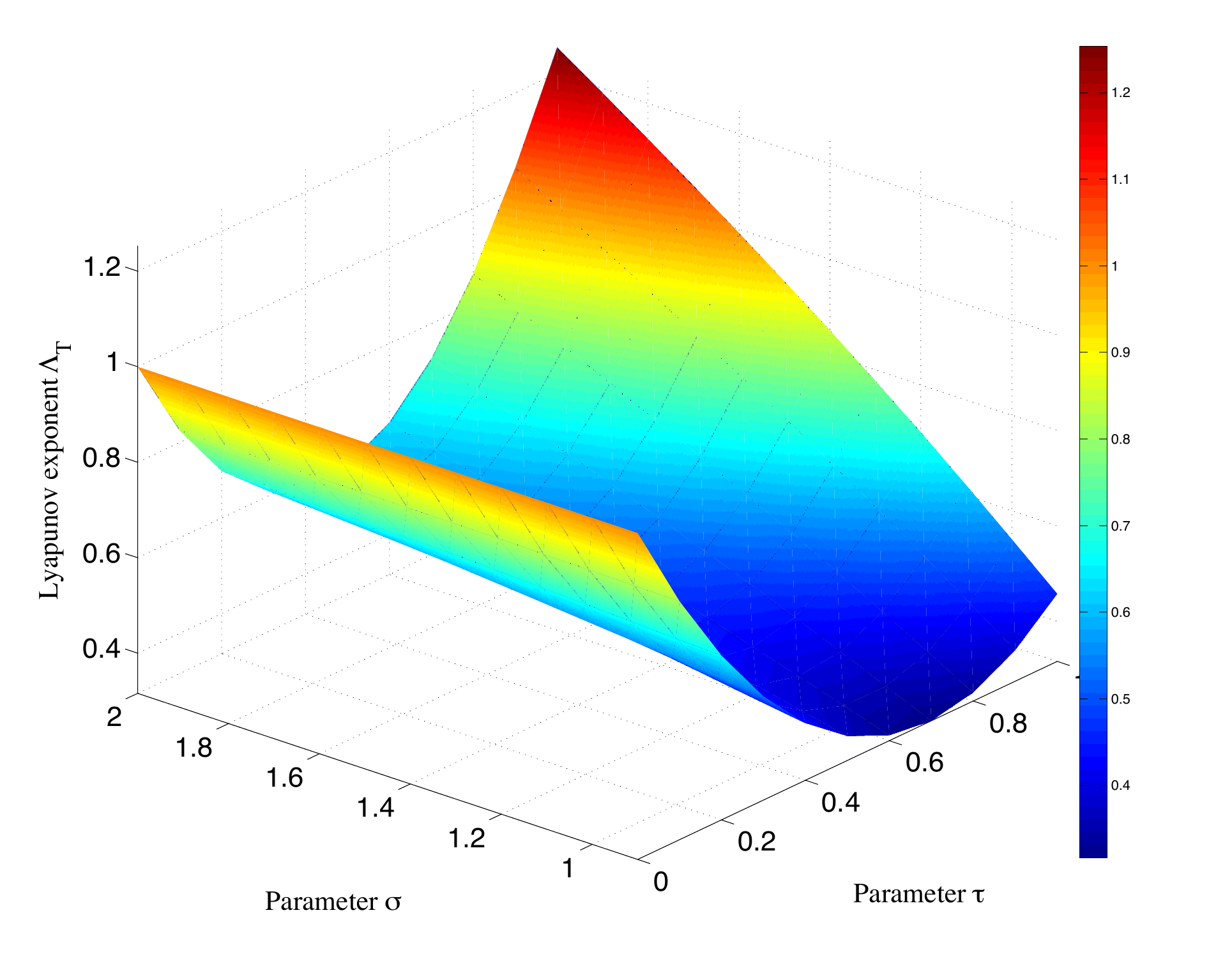}
\caption{Numerical estimation of the global largest Lyapunov exponent $\Lambda_T$ using algorithm \ref{alg: lambda(sigma)} as a function of $\tau$ and $\sigma$. \textbf{Left:} Input $u(t)=0$. \textbf{Right:} Input $u(t)=\sin(\omega t)$ with $\omega=0.25$. Other parameters: $T=1000$, $l=\alpha=m=1$. }
	\label{fig:lyapunov2}
\end{figure*}

In figure \ref{fig:lyapunov2}, we have applied algorithm \ref{alg: lambda(sigma)} to estimate $\Lambda_T$ in the case where $u(t)=0$ (left) and where $u(t) = \sin(\omega t)$ (right) for various values of parameters $\sigma$ and $\tau$. In this figure, one observes that $\Lambda_T$ is an increasing function of $\sigma$, which is a consequence of the fact that both $\gamma(t)$ and $\lambda(t)$ are increasing functions of $\sigma$, and corresponds with the intuition that increasing the disorder level would increase the unstability of the dynamics. The case of null input (left) with $\tau=1$ corresponds to the classical case \cite{PhysRevLett.69.3717}, and displays a kink at $\sigma=1$, whose consequences in terms of information processing has been discussed in \cite{toyoizumi2011beyond}. However, the impact of the leak rate $\tau$ on the Lyapunov exponent has not been studied so far to our knowledge, and reveals an interesting U-shaped behavior indicating that there exists an optimal intermediate value of $\tau$ which minimizes the instability of the system. Our purpose in the present paper is to evaluate the Lyapunov exponent when the system is driven by an external time-series, which is displayed on the right panel of figure \ref{fig:lyapunov2} with $u(t) = \sin(\omega t)$. This figure shows that the overall behavior is similar to the null-input case, with the expected difference that $\sigma$ must be set much larger than one (around 1.6 when $\tau=1$) to observe an exponent $\Lambda_T>1$. Intuitively, the driven system is more stable because the input acts as a time-dependent bias in the sigmoid transfer function, hence reducing its average slope $|S'|$ along a trajectory, and therefore the norm of the Jacobian matrix which controls the local expansion rate. Notice that the quantity $\Phi$ defined in \eqref{eq: phi} corresponds to the average squared slope of $S$, where the average will be taken with respect to the Gaussian distribution with appropriate time-dependent variance \eqref{eq:local}.

\section{Local Echo State Property}\label{sec: local ESP}

In this section, we discuss in more details the connection between the Lyapunov exponent $\Lambda_T$ and the ESP.

\subsection{Definition}
The intuition behind the ESP is that the network should follow a reproducible and robust attractor. If the attractor is not stable, then the output connectivity matrix would be learned on a trajectory which could be different from the trajectory observed during the prediction or test phase, leading to poor accuracy. A key element to quantify the stability of the network trajectory is to measure the impact of small perturbations. If these perturbations are amplified over time then the dynamics is too irregular for good performance, the network is chaotic. Therefore, we define a local version of the ESP which guarantees the robustness of the dynamics to perturbations:
\begin{defi}[Local ESP]
A driven dynamical system has the local Echo State Property if a small perturbation $\tilde{\x}(t_0) = \x(t_0) + \delta$ applied at time $t_0$ decreases to $0$ in the large time asymptotic limit, namely $||\x(t)-\tilde{\x}(t)||\to 0$ when $t\to \infty$ for $\delta$ sufficiently small. 
\end{defi}
This definition differs from the traditional ESP Definition \ref{def: ESP} in two aspects: first, it deals with perturbations which do no necessarily occur at time $t=-\infty$. This definition only asks the perturbed solution to converge eventually towards the unperturbed solution, whereas the traditional definition asks that the solutions are identical based on the fact that the perturbation occurred an infinite number of time steps before. This definition is closer to the practical application of ESN where the initial condition corresponds to $t=0$. Second, this definition only guarantees a local stability of the trajectories asking them to be robust only to small enough perturbations. On the other hand the traditional ESP requires that even large perturbations leave the trajectory unchanged. Put differently the traditional ESP guarantees a unique globally stable attractor, whereas the local ESP guarantees local stability of possibly many attractors (which have the same statistical properties).

We claim that the local ESP is sufficient for the good behavior of the network for most applications. More precisely, the only danger for systems that satisfy the local ESP, and not the traditional global ESP, is when learning is made on one attractor and prediction / test is made on another. In applications, if prediction / test is made immediately after learning such that we are sure to stay on the same attractor, then the local ESP is sufficient. On the other hand, if the initialization of the prediction / test phase is done randomly, then the network may converge to a different attractor than that explored during learning. In that case, one would expect the performance to be poor.

\subsection{Characterization}

Measuring the evolution of small perturbations precisely corresponds to computing the largest Lyapunov exponent. Indeed, if $\lambda < 1$ then a small perturbation will eventually vanish and the perturbed solution will converge to the unperturbed solution. Therefore, by construction we have the following quantitative criterion for the local ESP:
\begin{thm}
	If $\lambda[u] < 1$ then the network has the local ESP for the input $u$.
	\label{thm: lambda < 1}
\end{thm}

{Some remarks:}
\begin{itemize}
\item The local ESP can be valid for systems experiencing temporary growth of perturbations as long as they are followed by a more important decrease. What matters in the definition of the local ESP is the balance of growth and decrease over a long time.
\item From the key inequality \eqref{eq:inequality}, we deduce that the local ESP hold for all inputs whenever $\mu<1$. This is a further argument supporting the practical criterion of a spectral radius below 1 should work for all inputs.
\item There is a unique $\sigma_{L}$ such that $\lambda(\sigma=\sigma_L) = 1$ and that the local ESP holds for all $\sigma < \sigma_L$.
Indeed, we claim first that for any input $u$, the mapping $\sigma^2 \mapsto \lambda[u]$ is increasing. The proof is as follows. The function $F$ is increasing, concave with $F'(0) = 1$. Therefore, equation \eqref{eq: consistency variances} shows that $\gamma(t)$ increases sublinearly with $\sigma^2$. Performing a simple change of variable in equation \eqref{eq: phi}, it is easy to see that $\Phi(z^2)$ decreases slower that $1/z$ when $z^2$ increases. Therefore, $\sigma^2 \Phi\left(\alpha\sigma^2\gamma^2(t)+ m^2 u(t)^2 \right)$ increases with $\sigma^2$ and so does $\lambda_T$ according to equation \eqref{eq: lambda T}. Finally, one observes that $\lambda(\sigma=0) \leq 1$ and $\lambda(\sigma=+\infty) = + \infty$.
\end{itemize}

\subsection{Numerical experiments}
We now present an algorithm to compute $\lambda[u]$. A dichotomy algorithm, or any zero search algorithm for non-linear functions, could be implemented to find an approximation of $\sigma_L$, but given the cheap computational cost of computing $\lambda[u]$ for any $\sigma$, we will rather perform a grid search in this paper.

The algorithm to compute $\lambda_T[u]$ is stated below, when $\sigma, m, \alpha, l, \tau$ and $u$ have been fixed.
\begin{algorithm}
\caption{Computing $\lambda_T[u]$}
\begin{algorithmic}[1]
\State $\lambda \gets 1$
\State $\gamma^2 \gets 0$
\State $R, C, \gamma_{\text{hist}} \gets 0 \in \R^k$
\For{t = 1 : T}
\For{s = 1 : k-1} % s = 1 far away in the past
\State $C[s] \gets (1-l\tau) C[s+1] + \tau R[s+1]$
\State $R[s+1] \gets (1-l\tau) R[s] + \tau G(C[s],\gamma^2, \gamma_{\text{hist}}[s])$
\EndFor
\State $a \gets \alpha \sigma^2 \gamma^2 + m^2 u(t)^2$
\State $\lambda \gets \lambda \big((1 - l \tau)^2 + \tau^2 \alpha \sigma^2 \Phi(a) \big)^{1/T}$
\State $\gamma^2 \gets (1 - l \tau)^2 \gamma^2 + \tau^2 F(a) + 2\tau (1 - l \tau) R[-1]$
\State $\gamma_{\text{hist}}[:-1] \gets \gamma_{\text{hist}}[1:]$
\State $\gamma_{\text{hist}}[-1], C[k] \gets \gamma^2$
\EndFor
\State return $\lambda$
\end{algorithmic}
\label{alg: lambda(sigma)}
\end{algorithm}
Note that this algorithm is computationally cheap, in $O(T)$, especially compared to the simulation of the full network.
\begin{figure}[t]
\centering
\includegraphics[width=9 cm]{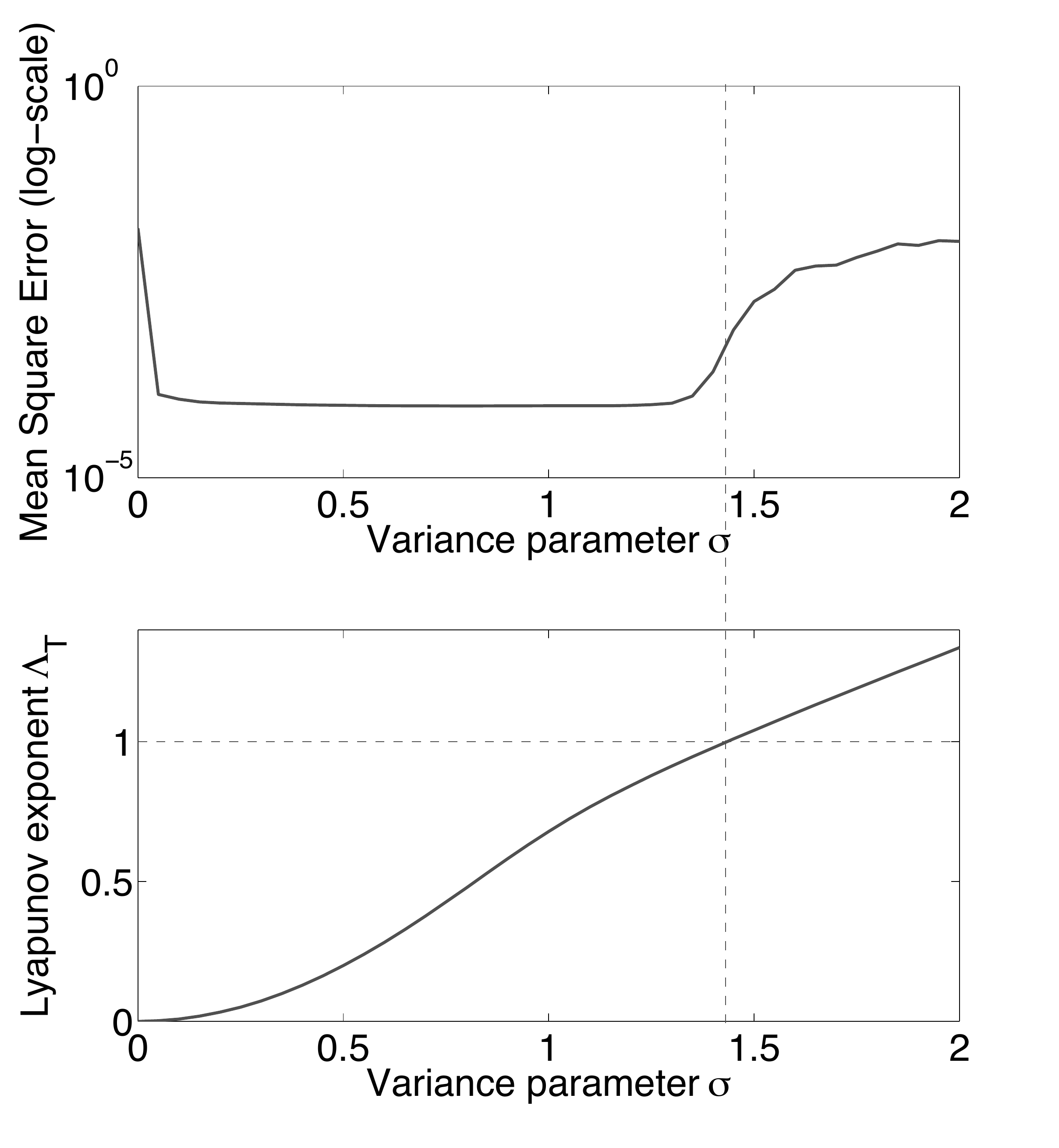}
\caption{\textbf{Top:} this figure displays the prediction accuracy (mean-square error on a testing set) of an ESN when $\sigma$ vary from $0$ to $2$. \textbf{Bottom:} this figure display the Lyapunov exponent $\Lambda_T$ as a function of $\sigma$. The dashed lines help seeing the critical value $\sigma^* \simeq 1.57$ which both corresponds to $\Lambda_T =1$ and to the transition to a regime of poor accuracy for the ESN. For the simulations the parameters were $n = 2000$, $\alpha = l = \tau = 1$, $m = 1.$ and $T = 2000$. The time-series to predict is a solution of the Mackey-Glass chaotic dynamical system with $\delta_{MG}=18$.}
\label{fig: input}
\end{figure}
\begin{figure}[t]
\centering
\includegraphics[height=7 cm]{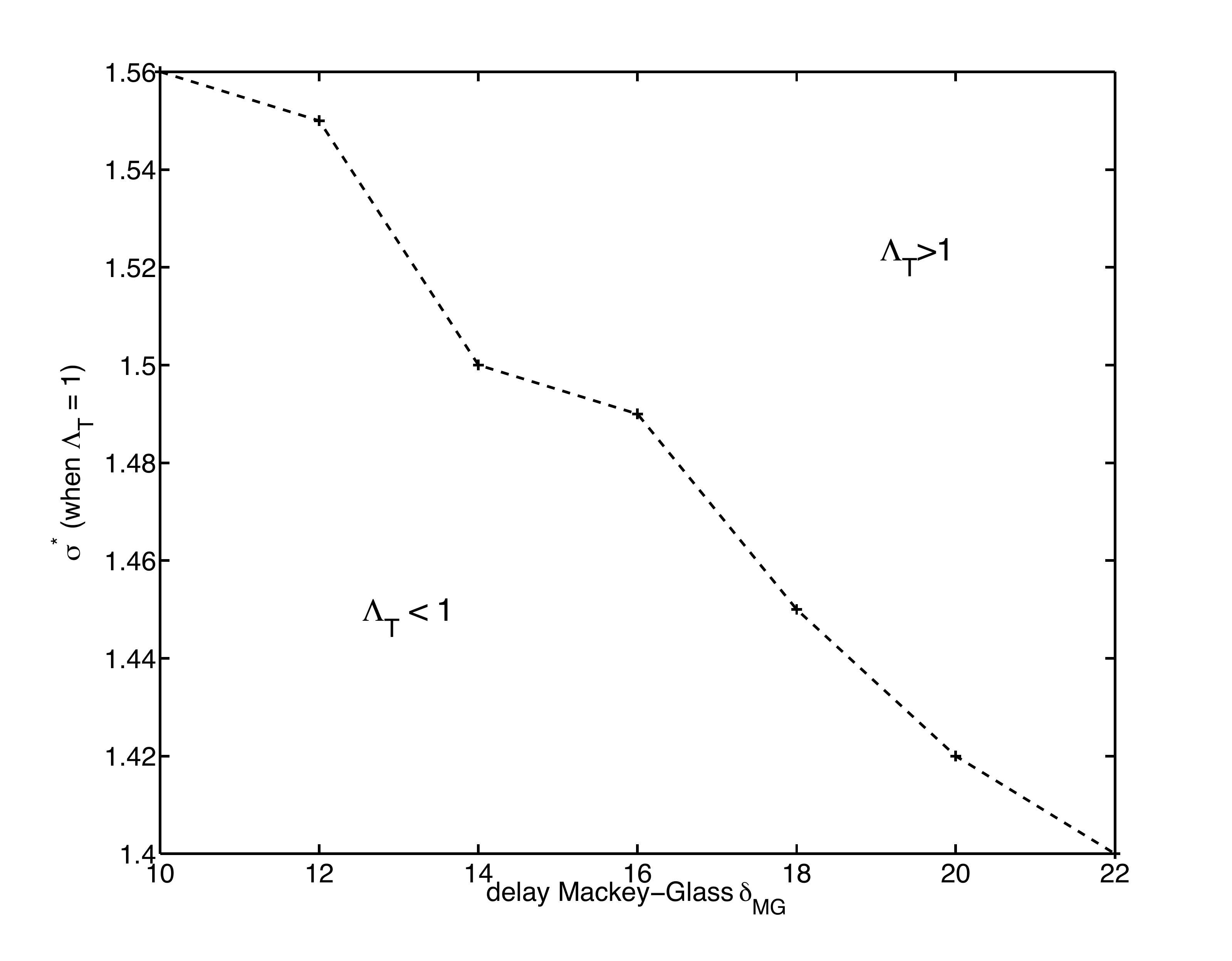}
\includegraphics[height=7 cm]{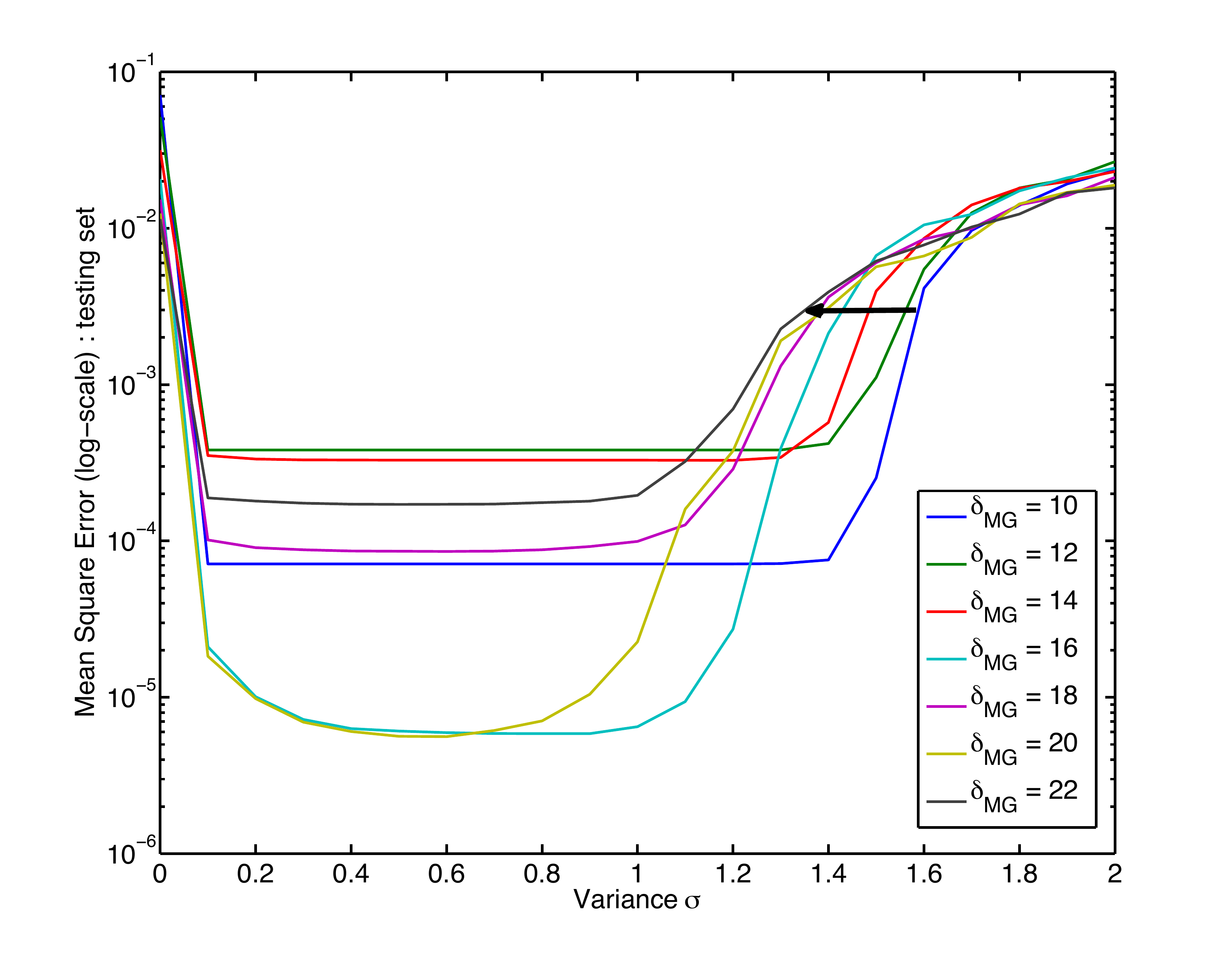}

\caption{Analysis of ESN performance as a function of the delay $\delta_{MG}$ in the Mackey-Glass prediction task. \textbf{Left:} Using Algorithm 1, we were able to compute the value $\sigma^*$ of the weights variance for which $\Lambda_T$ becomes larger than $1$, corresponding to the edge of chaos. The value of $\sigma^*$ depends on the delay parameter $\delta_{MG}$: it appears that increasing $\delta_{MG}$ leads to a smaller value of $\sigma^*$. \textbf{Right:} Mean-square error (testing set) as a function of the variance $\sigma$ for the Mackey-Glass prediction task, for different values of the delay $\delta_{MG}$. This figure confirms the prediction made in the Left panel : as indicated by the \textbf{black arrow}, for higher values of $\delta_{MG}$, the value of $\sigma$ where the performance starts to become poorer appears earlier.  For the simulations the parameters of the MG system were $a=0.2$, $b=0.1$ with a time-step $\delta t=1$ and the ESN parameters were $n = 100$, $\alpha = l = \tau = 1$, $m = 1$, for time-series of length $T = 2000$.}
\label{fig: mackeyglass}
\end{figure}

To show on a numerical example that the local ESP guarantees good accuracy, we have computed the prediction performance for a prediction task. More precisely, we consider here the classical task of Mackey-Glass (MG) time-series prediction. The MG dynamical system \cite{mackeyglass} is given by the following delayed differential equation:
\begin{equation}
	\dot{u}(t) = - b u(t) + \frac{a u(t-\delta_{MG})}{1+ u(t-\delta_{MG})^{10}}
\end{equation}
For each time-series, the task is to predict $u(t+1)$ (one-step ahead) given the past $u(1), ..., u(t-1),u(t)$. Training is done on half of the time-series and predictions are made for the other half. For different variances of the recurrent weights, we have plotted the accuracy of an ESN in figure \ref{fig: input} (top). This accuracy corresponds to the quantity $H = \frac{1}{T} \sum_{t=0}^{T-1} \big(u(t+1) - \w'.\x(t)\big)^2$, where $\w \in \R^n$ was computed with the usual Wiener-Hopf solution: $\w = \Big(\sum_{t=0}^{T-1} \x(t).\x(t)'\Big)^{-1}.\Big(\sum_{t=0}^{T-1} \x(t) u(t+1)\Big)$. We see that even for some $\sigma > 1$ the accuracy is good although the global ESP for all inputs is not satisfied any more. However, the accuracy becomes significantly poorer after a certain critical value for $\sigma$. In figure \ref{fig: input} (bottom), we have plotted the value of the Lyapunov exponent $\Lambda_T$ computed with the algorithm above. We see that it crosses $1$ quite precisely at a critical value $\sigma^*$ for which the accuracy moves to a regime of much higher values.

In order to further investigate the link between the local Lyapunov exponent $\Lambda_T$ and ESN performance, we have generated several discrete time-series corresponding to various values of $\delta_{MG} \in \{10,12,14,16,18,20,22\}$ with parameters $a=0.2$ and $b=0.1$.  In figure \ref{fig: mackeyglass} (right), ESN performance is measured by the Mean Square Error on a testing set and is displayed as a function of the variance parameter $\sigma$, for various values of the delay $\delta_{MG}$. Good performance is typically achieved for an intermediate range of values of $\sigma$, and one observes that the upper value of this range is smaller for higher values of $\delta_{MG}$, as indicated by the black arrow. We interpret this loss of performance for high values of $\sigma$ as related to the loss of the ESP. If this is indeed the case, then it should be possible to predict this behavior by using Algorithm 1 to compute $\sigma^*$, the value of $\sigma$ for which local Lyapunov exponent $\Lambda_T$ becomes larger than $1$. As displayed in figure \ref{fig: mackeyglass} (left), $\sigma^*$ is a decreasing function of $\delta_{MG}$, which is perfectly consistent with the above observation. This numerical example illustrates that the proposed theoretical advance presented in this article helps predicting and understanding the behavior of the performance curve for Echo-State Networks. However, finding the optimal value of all the hyper-parameters, beyond a systematic cross-validation procedure, remains a challenging theoretical problem.

\section{Conclusion}
In this paper, we have shown that the mean field theory for ESN developed in \cite{massar2013mean} can be, first, extended to leaky integrator networks on regular graphs; and, second, used to compute accurately a condition for the local ESP corresponding to the edge of chaos. We argue that the local ESP with respect to the given input is the useful condition to check in many applications, to ensure that the ESN representation is stable to small perturbations. We do not claim that the edge of chaos is always the best regime, but it has been shown that for some applications, typically requiring a lot of memory, it was optimal \cite{bertschinger2004real}. We believe that the proposed method to assess the local ESP should be systematically used to make sure the ESN has a regular dynamics leading to good accuracy. However, finding the optimal values of the hyper-parameters (e.g. $\sigma, \tau, \alpha$ etc.) for a given supervised learning task necessitates to take into account both the input and the target, which goes beyond the scope of the present approach : we provide a method to compute a bound for these parameters, given the input time-series, to ensure the ESP.

The theory has only been detailed for one dimensional inputs, but the extension of this approach to multidimensional inputs is not difficult (see \cite{massar2013mean}). Extending this method to other types of dynamics should be feasible as long as the computation of $F$ and $\Phi$ can be numerically done or conveniently reformulated. Finally, the mean-field approach only deals with the limit of very large networks $n\to \infty$, whereas in practice the aim might be to perform a given task with the smallest possible reservoir to avoid over-fitting issues. Therefore, a further investigation of the finite-size effects around the mean-field limit would be of interest. For instance, a related question has been studied in \cite{wainrib2013optimal}, where it is shown that networks with a variance parameter $\sigma<1$ have a probability to be unstable which is maximal for a specific size of the reservoir.

\section*{References}
\bibliographystyle{elsarticle-num}
\bibliography{./resa_proof_bib}

\end{document}